\documentclass[]{IEEEtran}
\usepackage{graphicx,color,amsmath,algorithmic,amsfonts,caption,subcaption}
\usepackage[caption=false]{subfig}

\begin{document}

\title{Gated Recurrent Unit (GRU) for Emotion Classification from Noisy Speech}
\author{\IEEEauthorblockN{Rajib Rana\IEEEauthorrefmark{1},
Julien Epps\IEEEauthorrefmark{2},
Raja Jurdak\IEEEauthorrefmark{3},
Xue Li\IEEEauthorrefmark{4}, 
Roland Goecke\IEEEauthorrefmark{5},
Margot Brereton\IEEEauthorrefmark{6}
and Jeffrey Soar\IEEEauthorrefmark{7}
%and
%Michael Breakspear\IEEEauthorrefmark{8}\\
}
\IEEEauthorblockA{\IEEEauthorrefmark{1}Institute of Resilient Regions (IRR)\\
University of Southern Queensland,
Springfield, QLD 4300\\ Email: rajib.rana@usq.edu.au}

\IEEEauthorblockA{\IEEEauthorrefmark{2}School of Electrical Engineering and Telecommunications, University of New South Wales\\
%j.epps@unsw.edu.au
}
\IEEEauthorblockA{\IEEEauthorrefmark{3}Distributed  Sensing Systems, CSIRO-Data61
%Email: raja.jurdak@csiro.au
}

\IEEEauthorblockA{\IEEEauthorrefmark{4}School of Information Technology and Electrical Engineering, University of Queensland\\
%xueli@itee.uq.edu.au
}
\IEEEauthorblockA{\IEEEauthorrefmark{5}Information Technology \& Engineering, University of Canberra\\
%roland.goecke@canberra.edu.au
}
\IEEEauthorblockA{\IEEEauthorrefmark{6}Computer Human Interaction, Queensland University of Technology\\
%jeffrey.soar@usq.edu.au
}
\IEEEauthorblockA{\IEEEauthorrefmark{7}School of Management and Enterprise, University of Southern Queensland\\
%jeffrey.soar@usq.edu.au
}
%\IEEEauthorblockA{\IEEEauthorrefmark{8}Systems Neuroscience Group, Queensland Institute of Medical Research\\
%Michael.Breakspear@qimrberghofer.edu.au
%}
}

\date{}
\maketitle

\begin{abstract}
Despite the enormous interest in emotion classification from speech, the impact of noise on emotion classification is not well understood. This is important because, due to the tremendous advancement of the smartphone technology, it can be a powerful medium for speech emotion recognition in  the outside laboratory natural environment, which is likely to incorporate background noise in the speech. We capitalize on the current breakthrough of Recurrent Neural Network (RNN) and seek to investigate its performance for emotion classification from noisy speech. We particularly focus on the recently proposed Gated Recurrent Unit (GRU), which is yet to be explored for emotion recognition from speech. Experiments conducted with speech compounded with eight different types of noises reveal that GRU incurs an 18.16\% smaller run-time while performing quite comparably to the Long Short-Term Memory (LSTM), which is the most popular Recurrent Neural Network proposed to date. This result is promising for any embedded platform in general and will initiate further studies to utilize GRU to its full potential for emotion recognition on smartphones.

%This research can open up opportunities for embedded mobile platforms, as GRU is  less computationally intensive but offers comparable accuracy to LSTM. Further large-scale experiments are warranted to strengthen the findings.
\end{abstract}

\section{Introduction}

Automatic Speech Emotion Recognition has gained increasing interest in both research and commercial space due to its tremendous potential to determine affective states~\cite{calix2012detection}.  
%Speech is a powerful modality to infer emotion and emotional responses are strong predictor of affective states. 
Automatic Speech emotion recognition can offer unprecedented opportunities for the Human-Computer Interaction Systems, as for example, recommendation systems can be designed to make more accurate affect-sensitive recommendations~\cite{shi2010mining}.
%; gaming consoles can make mood-aware adjustments in the game complexity and levels~\cite{schuller2015recent} and so on, thus making the games more interactive and interesting.  
This will also be highly beneficial for the burgeoning health and wellbeing applications as, for example, affect-diaries can be built to keep people aware of any prolonged negative mood, which would potentially prevent the onset or relapse of affective disorders~\cite{simon2003social}.

In recent years, Deep Learning models have revolutionized many fields, in particular, automatic speech recognition and computer vision~\cite{schmidhuber2015deep,DBLP:journals/corr/HannunCCCDEPSSCN14}. These models have also achieved improved performance in emotion recognition compared to conventional machine learning algorithms. For example, 
%deep hierarchical neural networks obtained the best reported results in detecting the Valence emotional dimension values and level of conflict~
\cite{brueckner2015odds,wollmer2010long,trigeorgis2016adieu,mao2014learning,li2013hybrid,wollmer2008abandoning,weninger2015introducing}, and 
%the use of autoencoders has improved unsupervised domain adaptation in affective speech analysis~
\cite{deng2014autoencoder} present some best reported results for affective speech analysis using deep learning models.

Out of the many forms of Deep Learning models, we consider Recurrent Neural Networks (RNNs) as they are targeted to model the long range dependencies between successive observations. Speech emotion possesses temporal dependency as it is unlikely to change rapidly between subsequent speech utterances~\cite{georgiev2014dsp}. Therefore, the Recurrent Neural Networks are most suitable for emotion recognition from speech. However, RNNs often suffer from the classical ``Vanishing'' and  ``Exploding'' Gradient problems, which results in failing to learn long-range dependencies. To avoid this, two variants of Recurrent Neural Networks have been proposed, which uses a ``gating'' approach to avoid these problems: (1) Long Short-Term Memory (1997)~\cite{hochreiter1997long} and (2) Gated Recurrent Unit (2014)~\cite{cho2014properties}. 

Smartphones are great platforms for speech emotion recognition, as people are close to their phones for an extended period of time. Research shows that almost 79\% of people have their smartphones with them 22 hours a day\footnote{http://www.adweek.com/socialtimes/smartphones/480485}. In addition, the processing capacity of the modern smartphones is also extraordinary.
% In fact they are considered desktop-grade. 
In the scope of this paper we assume that speech samples are collected during phone conversation~\cite{rana2016poster}. 

%One possible way to infer emotional states is during phone conversation. 
%
%Mood and emotion although are used interchangeably, mood is different from emotion. Emotion is rather transient but mood is long lasting. But we previously previously proposed that mood can be determined by analyzing the emotional response during phone conversation. In this paper we therefore focus on emotion recognition from speech. 
%
%
%Our end goal is to run the mood inference system on the smartphone. 
The key challenge of emotion recognition from phone conversation is background noise as people can talk over phone anywhere including the cafeteria, park, office and many more places. The mainstream research in emotion recognition from speech has mainly focused on clean speech captured in controlled laboratory environment. Therefore, the impact of noise in emotion classification is not well understood~\cite{yang2014does}.

Another major hurdle of running any Deep Learning models on a smartphone is computational complexity.  Despite the advancement in the processing capacity, smartphones are still limited by battery power. 
%Due to this, despite tremendous potentials speech are usually avoided to infer emotion on smartphones~\cite{likamwa2013moodscope}. 
Therefore, a method with a small runtime is a must.  Amongst LSTM and GRU, GRU has relatively simplified architecture, therefore, our focus is mainly on GRU. A number of studies have looked into the performance of LSTM (e.g., \cite{wei2014multimodal,wollmer2013lstm,tian2015emotion}) for emotion recognition from speech, however, the  performance of GRU for that is currently unexplored.

In this paper, we address the above two challenges. The contributions of this paper are as follows:
\begin{enumerate}
\item To the best of our knowledge we for the first time analyze the accuracy and the run-time complexity of Gated Recurrent Unit for emotion classification from speech. 
\item We superimpose various real-life noises on clean speech and analyze the classification performance of the Gated Recurrent Unit under various noise exposures.
\item We use LSTM as the benchmark and conduct a detailed accuracy and run-time comparisons of these two gating Recurrent units for emotion classification from noisy speech.
\end{enumerate}

The paper is organized as follows. In next section, we provide background information on Recurrent Neural Networks followed by the description of how GRU can be used for emotion classification from a noisy speech in Section~\ref{sec:GRU}. We then present the results, and discussion in Section~\ref{sec:ER}. This is followed by existing  work in Section~\ref{sec:EW}; and finally, we conclude in Section~\ref{sec:CON}.
\begin{figure}[h]
\centering
\includegraphics[width = 0.8\linewidth]{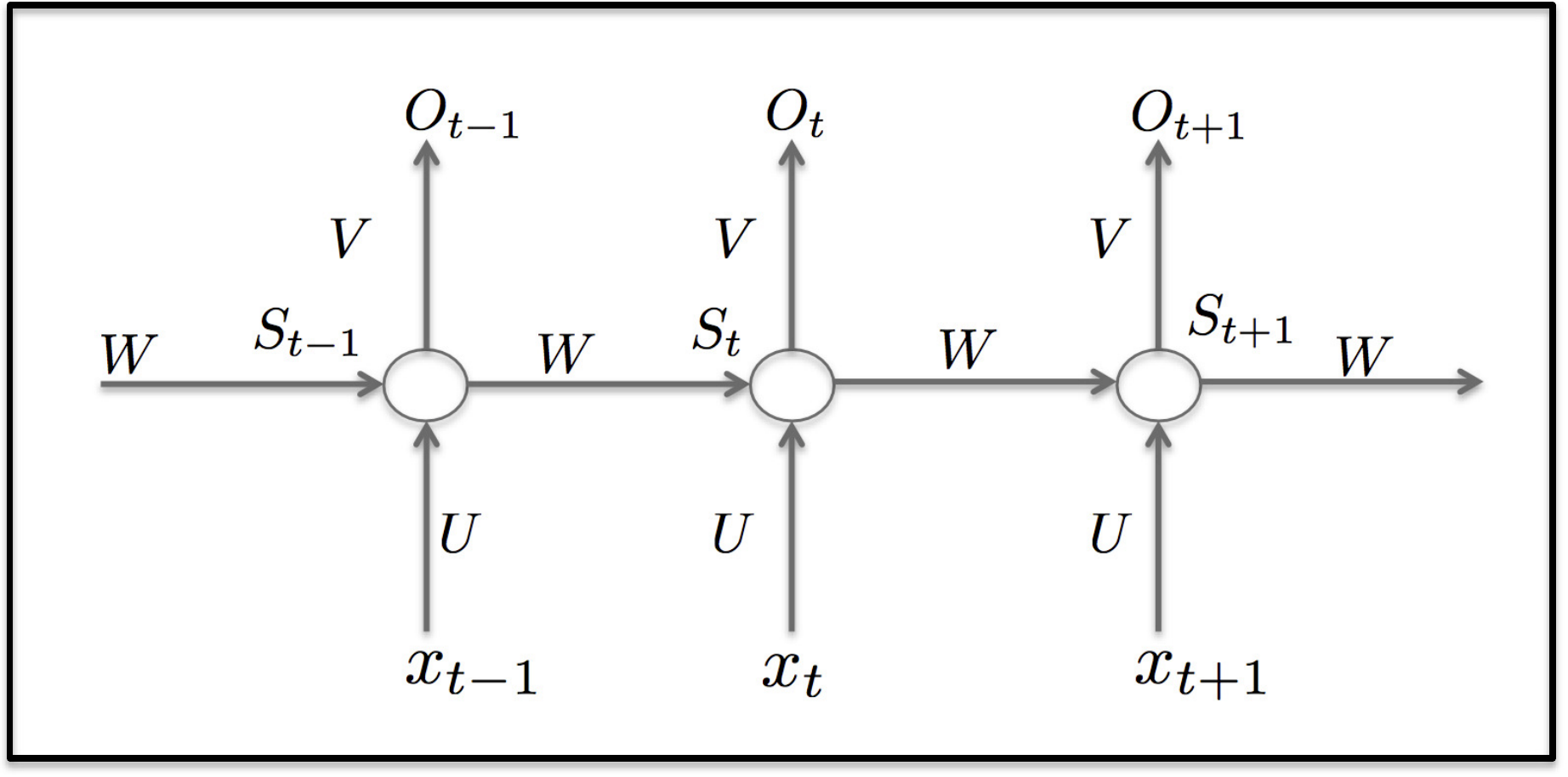}
\caption{Recurrent Neural Networks. 
%s are ; are ; s areMatrix $maps the $d$-dimensional input to a $p$ dimensional hidden unit; Matrix $V\in \mathbb{R}^{p\times \prime{d}}$ maps the $p$ dimensional hidden unit to output; Matrix $W\in \mathbb{R}^{p\times p}$ maps the $p$ dimensional hidden unit to another $p$ dimensional hidden unit.
}
\label{fig:rnn}
\end{figure}

\section{Recurrent Neural Networks}
\label{sec:RNN}
%\subsection{Recurrent Neural Networks} 
The conventional Feed Forward Neural Network is trained on labeled data until it minimizes the prediction error. Whereas the decision an RNN reached at time step $t-1$ affects the decision it will reach at time step $t$. Therefore, RNNs have two input sources: the present, and the recent past, which it uses in combination in determining a response to a new input.

An example Recurrent Neural Networks (RNNs) is shown in Figure~\ref{fig:rnn} and the related symbols are defined in Table~\ref{tab:symbol}.  A hidden state ($s_t$) is a non-linear transformation of a linear combination of input ($x_t$) and previous hidden state ($s_{t-1}$). Output ($o_t$) at time $t$ is only dependent on the current hidden state $s_t$. The output is associated with a probability $p_t$ determined through a ``softmax'' function. In Neural Networks a softmax function is implemented in the final layer of a network used for classification. For a classification with $K$ classes, the softmax function determines the probability of a probe being classified as each of the $K$ classes.
\begin{eqnarray}
%a_t &=&  Ws_{t-1} + Ux_t  \nonumber\\
% s_t&=& tanh(a_t)\nonumber\\
 s_t &=& tanh(Ws_{t-1} + Ux_t)\nonumber\\
o_t &=& Vs_t   \nonumber
%p_t &=& softmax(o_t)\nonumber 
\end{eqnarray}

\begin{table}[ht]
\centering
\caption{Symbol Definitions}
\begin{tabular}{|c|c|}\hline
{\bf Symbol} & {\bf Definition}\\\hline
 $x_t \in \mathbb{R}^d$ &$d$ dimensional input vectors \\ \hline
 $S_t \in \mathbb{R}^p$ & $p$ dimensional hidden unit\\ \hline
$O_t\in \mathbb{R}^{d^\prime}$& $d^\prime$ dimensional outputs\\ \hline
$U\in \mathbb{R}^{d\times p}$ & maps $d$-dimensional input to $p$ dimensional hidden unit\\ \hline
$V\in \mathbb{R}^{p\times d^ \prime}$ & maps $p$ dimensional hidden unit to $d^ \prime$ dimensional output\\ \hline
$W\in \mathbb{R}^{p\times p}$  & maps $p$ dimensional hidden unit to another hidden unit.\\ \hline
\end{tabular}
\label{tab:symbol}
\end{table}

\noindent\emph{Vanishing and Exploding Gradients:}
%Recurrent nets seeking to establish connections between a final output and events many time steps before were vanished, because it is very difficult to know how much importance to accord to remote inputs.  
Both of these terms were introduced by Bengio et al.~\cite{bengio1994learning} in 1994. The exploding gradient problem refers to the large increase in the norm of the gradient during training. The vanishing gradient problem refers to the opposite behavior when long-term components go exponentially fast to norm 0, making it impossible for the model to learn the correlation between temporally distant events.

Any quantity multiplied frequently by an amount slightly greater than one can become immeasurably large (exploding).  Multiplying by a quantity less than one is also true (vanishing). Because the layers and time steps of deep neural networks relate to each other through multiplication, derivatives are susceptible to vanishing or exploding.

%There have been two dominant approaches by which many researchers have tried to reduce the negative impacts of this issue. One such approach is to devise a better learning algorithm than a simple stochastic gradient descent [see, e.g., Bengio et al., 2013, Pascanu et al., 2013, Martens and Sutskever, 2011], for example using the very simple clipped gradient, by which the norm of the gradient vector is clipped, or using second-order methods which may be less sensitive to the issue if the second derivatives follow the same growth pattern as the first derivatives (which is not guaranteed to be the case).
%The other approach, in which we are more interested in this paper, is to design a more sophisticated activation function than a usual activation function, consisting of affine transformation followed by a simple element-wise nonlinearity by using gating units. The earliest attempt in this direction resulted in an activation function, or a recurrent unit, called a long short-term memory (LSTM) unit [Hochreiter and Schmidhuber, 1997]. More recently, another type of recurrent unit, to which we refer as a gated recurrent unit (GRU), was proposed by Cho et al. [2014]. 

Encouragingly, there are a few ways to address these gradient problems. Proper initialization of the weight (W) matrix and regularization can reduce the impact of vanishing gradients. Another possible solution is to use the Rectified Linear Unit (ReLU)~\cite{nair2010rectified} instead of the conventional $tanh$ or sigmoid activation functions. The ReLU derivative is either $0$ or $1$, so it is not as likely to cause a  vanishing or exploding gradient. However, ReLU units can be fragile during training and can ``die''. For example, a large gradient flowing through a ReLU neuron could cause the weights to update in such a way that the neuron will never activate on any data point again. As a result as much as 40\% of the neurons will never be activated across the entire training dataset if the learning rate is set too high. With a  rigorously chosen learning rate, this can be avoided. 

An even more effective solution is to use the gated Recurrent Neural Network architectures: Long Short-Term Memory (LSTM) or Gated Recurrent Unit (GRU). LSTMs first proposed by Hochreiter et al. in 1997 are one of the most widely used Recurrent Neural Networks today. GRUs, first proposed in 2014, are simplified versions of LSTMs. Both of these RNN architectures were purposely built to address the vanishing and the exploding gradient problem and efficiently learn long-range dependencies. To assist the understanding of GRU in the next section we first describe LSTM. 

%GRU has been built upon LSTM  and in fact it is a simplified version of LSTM. We therefore describe LSTM first which will assist in describing the GRU architecture. 
%This is where the ``gating'' was proposed. 

\begin{figure*}[ht]
\centering
   \begin{subfigure}{.45\columnwidth}
        \includegraphics[scale=.55]{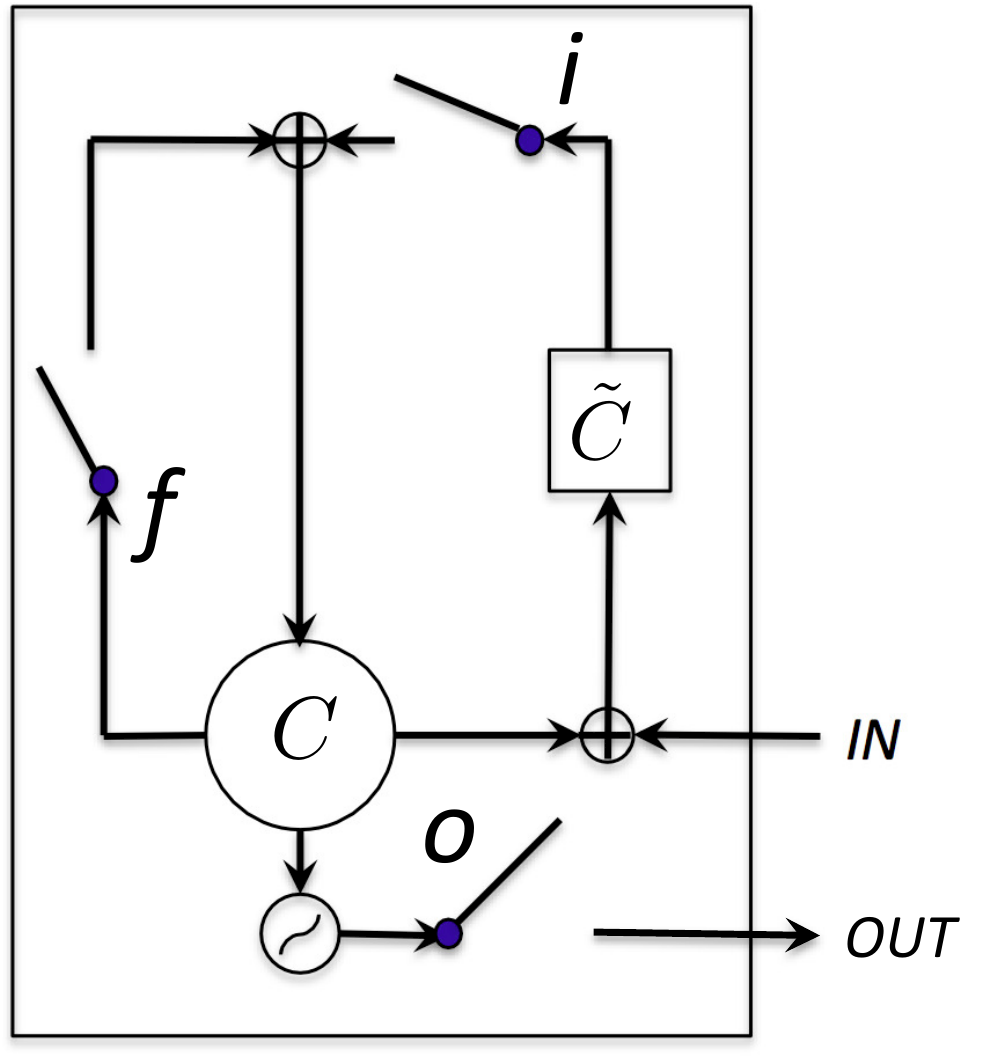}
       \caption{Long Short Term Memory Unit. \\$i$: Input Gate, $f$: Forget Gate, $o$: Output Gate.}
	\label{fig:lstmnew}
    \end{subfigure}
\hspace{2.5cm}
    \begin{subfigure}{.45\columnwidth}
        \includegraphics[scale=.55]{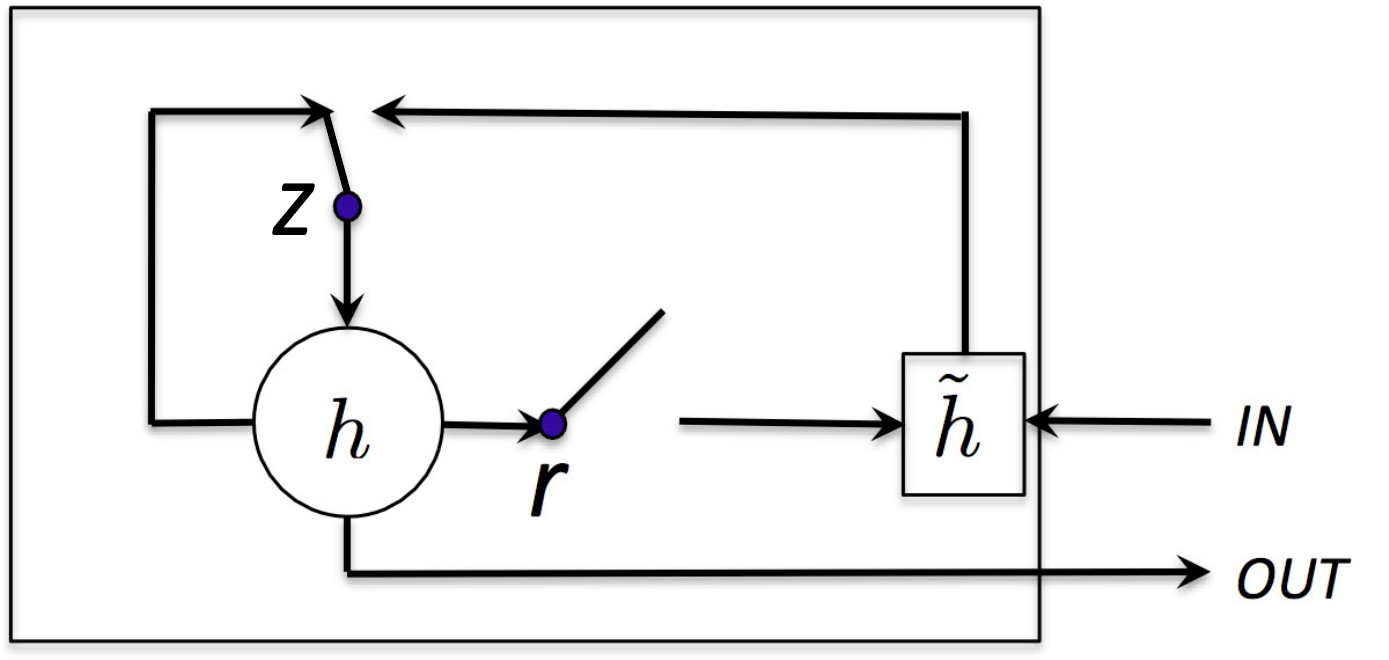}
        \caption{Gated Recurrent Unit. \\$z$: update Gate, $r$: Reset Gate.}
	\label{fig:gru}
    \end{subfigure}
\caption{Gated Recurrent Neural Networks.}
\end{figure*}

\subsection{Long Short-Term Memory Units (LSTMs)}
The Long Short Term memory is a special kind of Recurrent Neural Networks, that eliminates the shortcoming of vanishing or exploding gradient problem of the Recurrent Neural Networks. This makes LSTM suitable to learn from history to classify, process and predict time series when there are very long and unknown time lags between important events. An LSTM network is made up of LSTM blocks that have three gates: \emph{input}, \emph{output} and \emph{forget} gate, which help it to remember a value for an arbitrary long time; forget the value when it is not important to remember anymore or output the value. 
LSTMs help preserve the error that can be backpropagated through time and layers. By maintaining a more constant error, they allow recurrent nets to continue to learn over many time steps, for example over 1000.
%, thereby opening a channel to link causes and effects remotely.

LSTMs contain information outside the normal flow of the recurrent network in a gated cell. Information can be stored in, written to, or read from a cell, much like data in a computer's memory. The cell makes decisions about what to store, and when to allow reads, writes, and erasures, via gates that open and close. Unlike the digital storage on computers, however, these gates are analog, implemented with element-wise multiplication by sigmoids, which are all in the range of $[0 - 1]$. Analog has the advantage over digital of being differentiable, and therefore suitable for backpropagation.

Those gates act on the signals they receive, and similar to the neural network’s nodes, they block or pass on information based on its strength and import, which they filter with their own sets of weights. Those weights, like the weights that modulate input and hidden states, are adjusted via the recurrent networks learning process. That is, the cells learn when to allow data to enter, leave or be deleted through the iterative process of making predictions, backpropagating error, and adjusting weights via gradient descent.

LSTMs’ memory cells give different roles to addition and multiplication in the transformation of input. Instead of determining the subsequent cell state by multiplying its current state with new input, they add the two, which helps them preserve a constant error when it must be backpropagated at depth. 

%The central plus sign in both diagrams is essentially the secret of LSTMs. Stupidly simple as it may seem, this basic change helps them preserve a constant error when it must be backpropagated at depth. Instead of determining the subsequent cell state by multiplying its current state with new input, they add the two, and that quite literally makes the difference. (The forget gate still relies on multiplication, of course.)

Different sets of weights filter the input for input, output, and forgetting. The forget gate is represented as a linear identity function, because if the gate is open, the current state of the memory cell is simply multiplied by one, to propagate forward one more time step.

The LSTMs need a forget gate although their purpose is to link distant occurrences to a final output. This can be justified with an example. While analyzing a text corpus when the pointer comes to the end of a document, it could be the case the next document does not have a correlation with the previous one. Therefore, the memory cell should be set to zero before the net ingests the first element of the next document.
 
An LSTM cell has been shown in Fig~\ref{fig:lstmnew}. It is not limited to computing the weighted sum of the inputs and then appliying a nonlinear function; rather each $j$-th LSTM unit maintains a memory $c_t^j$ at time $t$. The activation function of the LSTM is 
\begin{eqnarray}
h_t^j = o_t^j \tanh{c_t^j}\nonumber.
\end{eqnarray}
The output gate $o_t^j$ modulates the amount of memory content exposure. With $V_o$ as a diagonal matrix, It is calculated by 
\begin{eqnarray}
o_t^j = \sigma (W_o x_t + U_f h_{t-1} + V_o c_t)^j\nonumber.
\end{eqnarray}
The memory $c_t^j$  is updated by partially forgetting the existing memory and adding a new memory $\tilde{c}_t^j$. The extent to which the existing memory is forgotten is controlled by a forget gate $f_t^j$. 
\begin{eqnarray}
f_t^j = \sigma (W_f x_t + U_o h_{t-1} + V_i c_{t-1})^j\nonumber.
\end{eqnarray}
And the extent to which new memory is added is controlled by an input gate $i_t^j$
\begin{eqnarray}
i_t^j = \sigma (W_i x_t + U_i h_{t-1} + V_f c_{t-1})^j\nonumber.
\end{eqnarray}
Controlled by these gates the existing memory is updated using the following equations. 
\begin{eqnarray}
\tilde{c}_t^j &=&\tanh{(W_c x_t + U_c h_{t-1})^j},\nonumber\\
{c}_t^j &=&f_t^jc_{t-1}^j + i_t^j\tilde{c}_t^j.\nonumber
\end{eqnarray}

\begin{figure*}[t]
\centering
\includegraphics[width =1\linewidth]{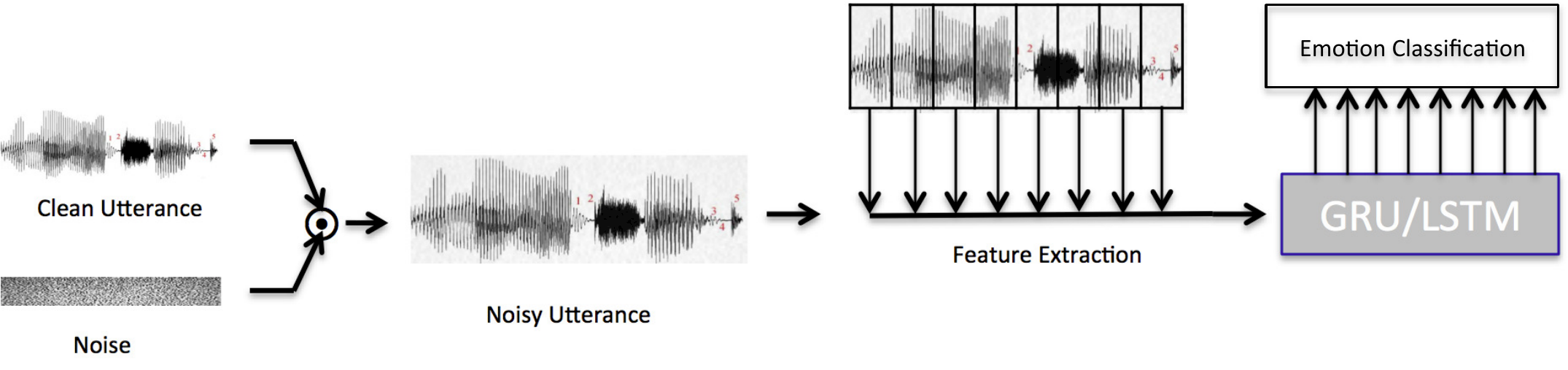}
\caption{Experimental setting for emotion classification from noisy speech.}
\label{fig:noisyFramework}
\end{figure*}
\begin{table*}[ht]
\centering
\caption{LSTM vs GRU}
\begin{tabular}{|c|c|c|}\hline
&LSTM&GRU \\ \hline
{\bf Controlled Memory Exposure} &The amount of memory seen by the other & The whole memory is exposed to the network\\
 &units of the network is controlled by the Output Gate &\\ \hline
{\bf New Memory Computation} &No separate control for amount of & Controls the information flow from the previous activation \\
&information flow from the previous time step&\\ \hline
{\bf Complexity vs Performance}&  With an additional gate &  Has fewer parameters and\\
 &is likely to have higher complexity &thus may train comparatively faster or need less data to generalize \\ \hline
\end{tabular}
\label{tab:gruVSlstm}
\end{table*}

\subsection{Gated Recurrent Units (GRUs)}
The Gated Recurrent Unit is a slightly more simplified variation of the LSTM. It combines the forget and input gates into a single ``update gate'' and has an additional ``reset gate''. The end model is simpler than standard LSTM models and is becoming increasingly popular.

A Gated Recurrent Unit like LSTM modulates information inside the unit, however, without having a separate memory cell (see Fig~\ref{fig:gru}). The activation $h_t^j$ of the GRU at time $t$ is a linear interpolation between the previous activation $h_{t-1}^j$ and the candidate activation $\tilde{h}_t^j:$
\begin{eqnarray}
h_t^j = (1 - z_t^j)h_{t-1}^j + z_t^j\tilde{h}_t^j\nonumber
\end{eqnarray}
The update gate $z_t^j$ decides how much the unit updates its activation.
\begin{eqnarray}
z_t^j =\sigma(W_zx_t + U_z h_{t-1})^j.\nonumber
\end{eqnarray}
The candidate activation $\tilde{h}_t^j$ is computed similarly to the update gate:
\begin{eqnarray}
\tilde{h}_t^j = tanh(Wx_t + U (r_t .* h_{t-1}))^j,\nonumber
\end{eqnarray}
\cite{hochreiter1997long}
where $r_t^j$ is a set of reset gates and $.*$ denotes an element-wise multiplication. When the reset gate is off ($r_t^j == 0$), it allows the unit to forget the past. This is analogous to letting the unit reading the first symbol of an input sequence. The reset gate is computed using the following formula 
\begin{eqnarray}
r_t^j =  \sigma(W_rx_t + _rh_{t-1})^j\nonumber
\end{eqnarray}

%If update gate is around 0, previous memory is ignored, and only new information is stored. The reset gate controls whether the input or the previous state determines the current state. If reset is close to 0, ignore previous hidden state. Allows model to drop information that is irrelevant in the future. 
Update gate $z$ controls how much the past state should matter now. Units with short-term dependencies will have active reset gates $r$. Units with long-term dependencies have active update gates $z$.

GRU and LSTM have many commonalities, yet there are some differences between these two, which we summarize in Table~\ref{tab:gruVSlstm}.

\section{Gated Recurrent Unit for Emotion Classification from Noisy Speech}
\label{sec:GRU}
An emotion classification framework embodying a Gated  Recurrent Unit is shown in Fig.~\ref{fig:noisyFramework}.  Our focus is on emotion classification from noisy speech, so we simulate noisy speech upon superimposing various environmental noises on clean speech. Features are extracted from the noisy speech and feed to the GRU for emotion classification. We have used the same framework for LSTM to contrast its classification performance with that of GRU.
%Utterances are divided into segments and passed onto the networks for segment level emotion classification. We will adopt method proposed in~\cite{han2014speech} to determine utterance level classification from the segment.

\subsection{Description of Datasets}
The Berlin emotional speech database~\cite{burkhardt2005database} is used in experiments for classifying discrete emotions. In this database, ten actors, five males and five females each uttered ten sentences (5 short and 5 longer, typically between 1.5 and 4 seconds) in German to simulate seven different emotions: anger, boredom, disgust, anxiety/fear, happiness, and sadness. Utterances scoring higher than 80\% emotion recognition rate in a subjective listening test are included in the database. We classify all the seven emotions in this work. The numbers of speech files for these emotion categories in the presented Berlin database are: anger (127), boredom (81), disgust (46), fear (69), joy (71), neutral (79) and sadness (62). We use this prototypical database for the preliminary investigation of GRU's performance for emotion recognition. 

In order to simulate noise-corrupted speech signals, the DEMAND (Diverse Environments Multi-channel Acoustic Noise Database) noise database~\cite{thiemann2013diverse} has been used in this paper. This database involves 18 types of noises wherein we have used noise from traffic, cafe, living room, park, washing, car, office and river. The recordings were captured at a sampling rate of 48 kHz and with a target length of 5 minutes (300 s). Actual audio capture time was slightly longer thereby allowing the removal of set-up noises and other artifacts by trimming. The recorded signals were not subject to any gain normalization. Therefore, the original noise power in each environment was preserved.
\begin{figure*}[ht]
\centering
   \begin{subfigure}{.45\linewidth}
	\centering
        \includegraphics[scale=0.4]{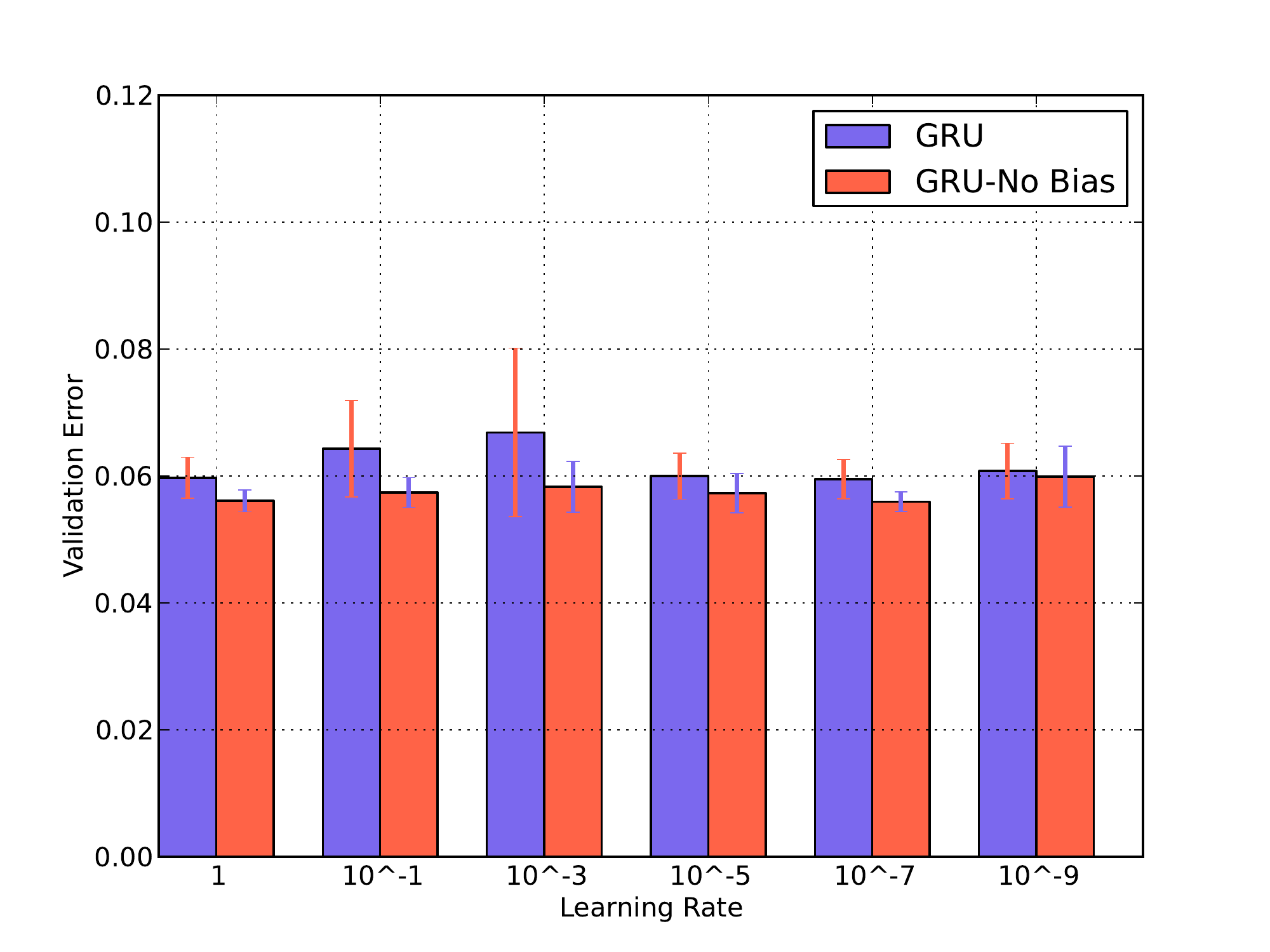}
         \caption{GRU - No Noise.}
	\label{fig:BiasLRateGRU}
    \end{subfigure}
   \begin{subfigure}{.45\linewidth}
	\centering
        \includegraphics[scale=0.4]{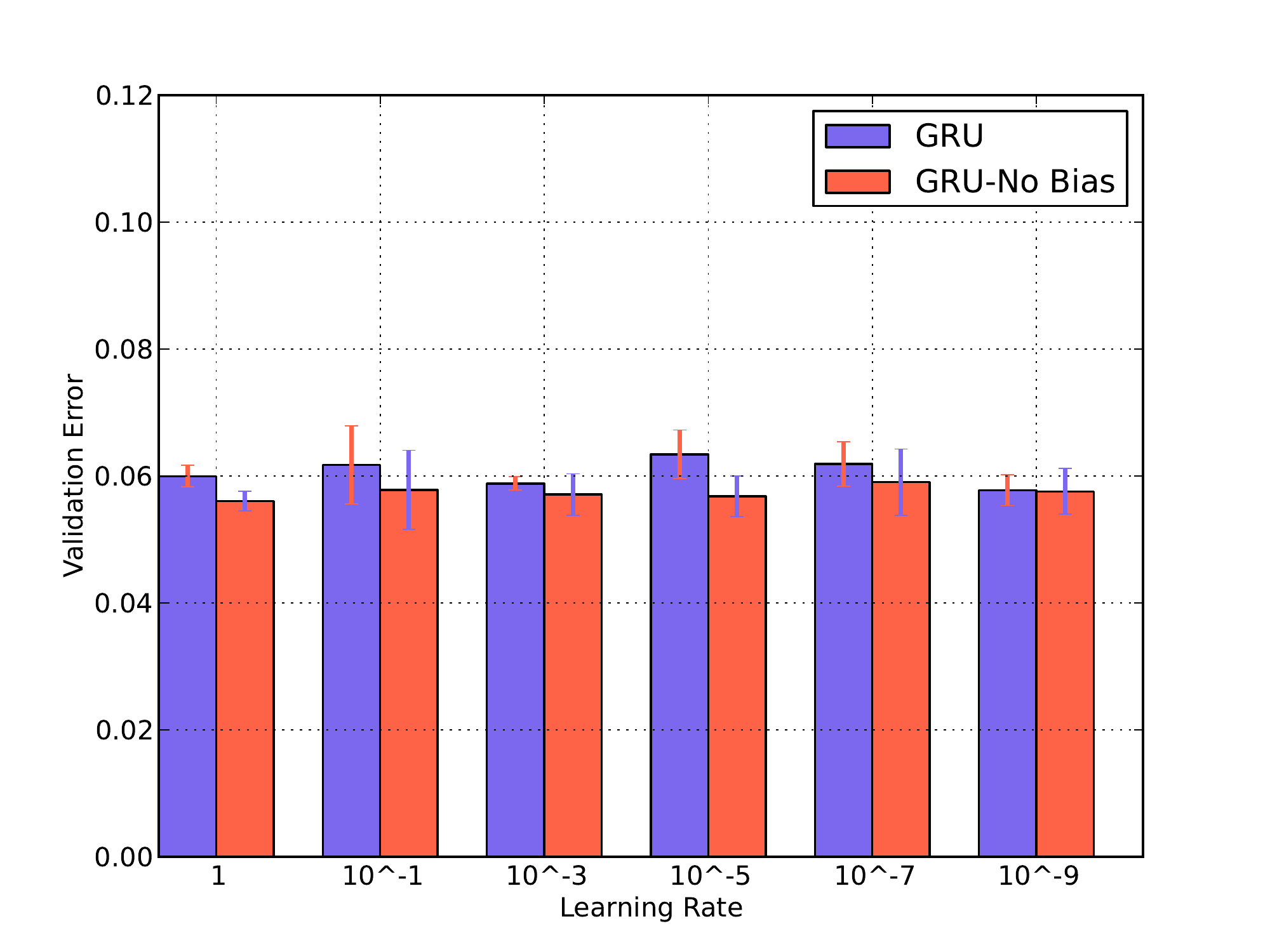}
         \caption{GRU - Traffic Noise.}
	\label{fig:BiasLRateGRU}
    \end{subfigure}
    \begin{subfigure}{.45\linewidth}
	\centering
        \includegraphics[scale=0.4]{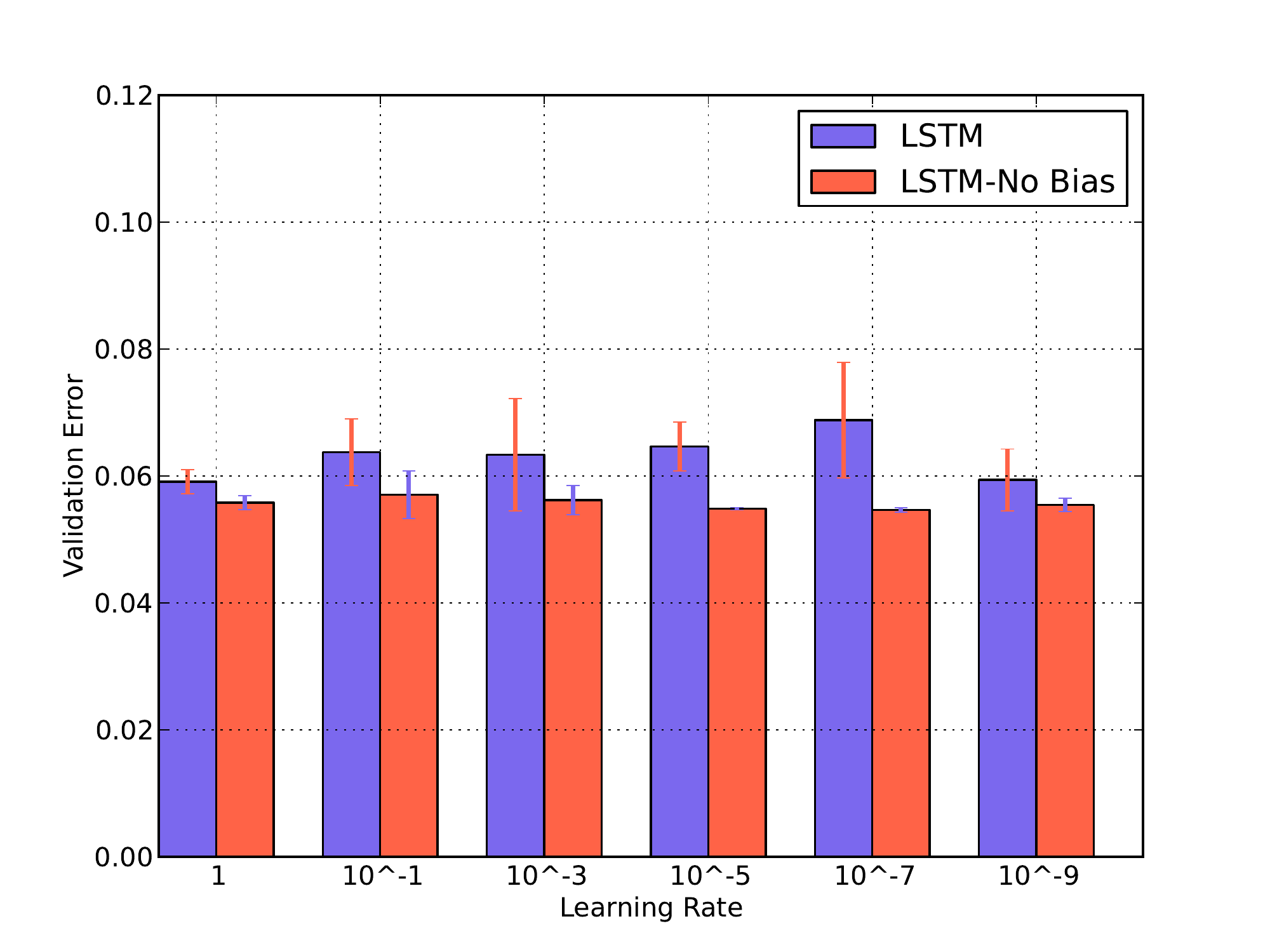}
         \caption{LSTM - No Noise.}
	\label{fig:BiasLRateGRU}
    \end{subfigure}
    \begin{subfigure}{.45\linewidth}
	\centering
        \includegraphics[scale=0.4]{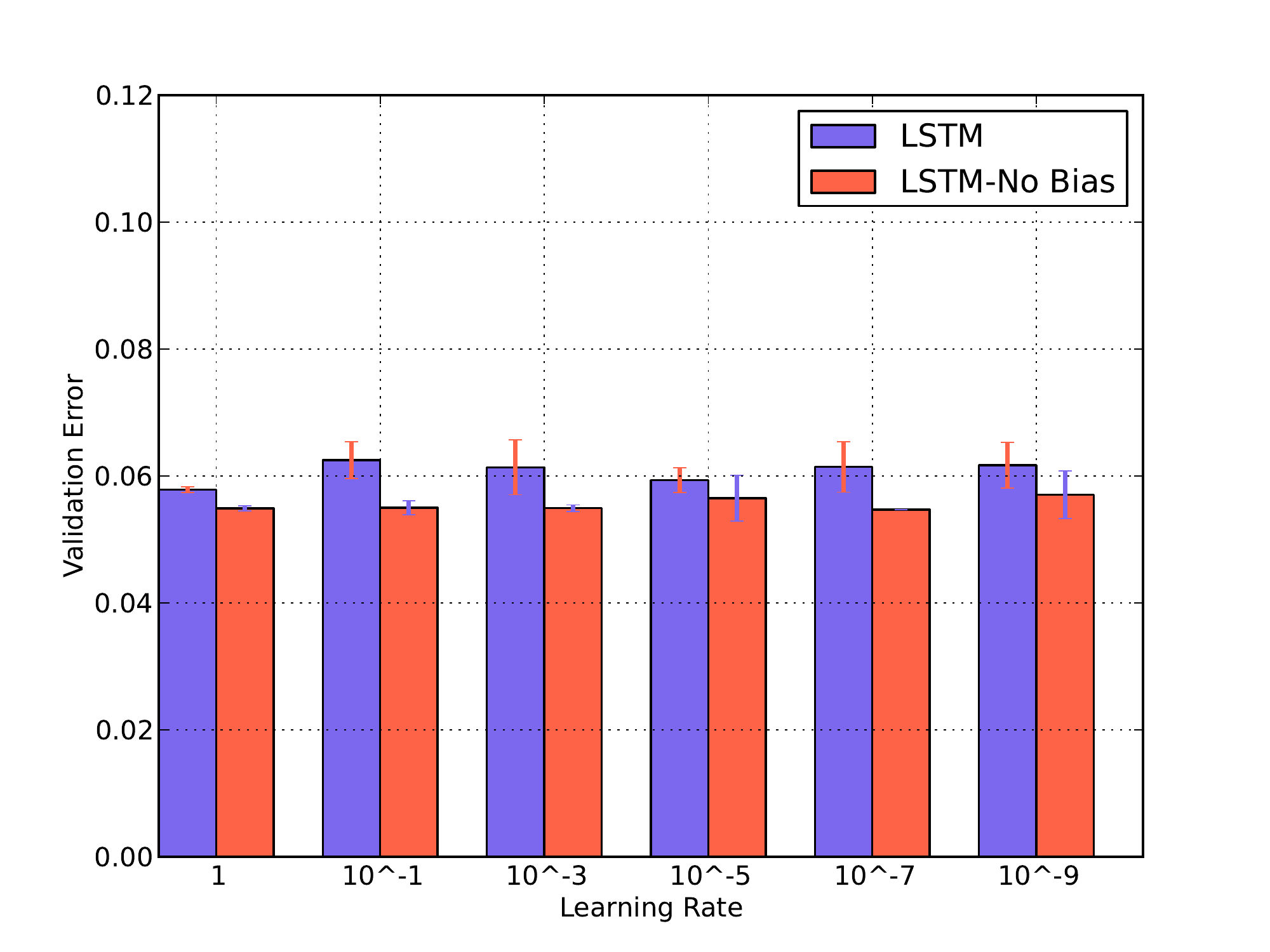}
         \caption{LSTM - Traffic Noise.}
	\label{fig:BiasLRateGRU}
    \end{subfigure}
\caption{Classification Performance - Impact of Noise, Bias and Learning Rate.}
\label{fig:cpiNLB}
\end{figure*}

\subsection{GRU Implementation}
We use the PyBrain Toolbox~\cite{schaul2010pybrain} 
to implement the GRU. We evaluate the classification performance across three parameters including \emph{learning rate}, \emph{number of cells} and \emph{bias}.  To use GRU and LSTM for classification we  use softmax function in the output layer. The Pybrain package uses $75\%$ of data for training and $25\%$ for validation. For accuracy we use the validation error in the plots.

\begin{figure*}[ht]
\centering
\begin{subfigure}{.45\linewidth}
	\centering
        \includegraphics[scale=0.4]{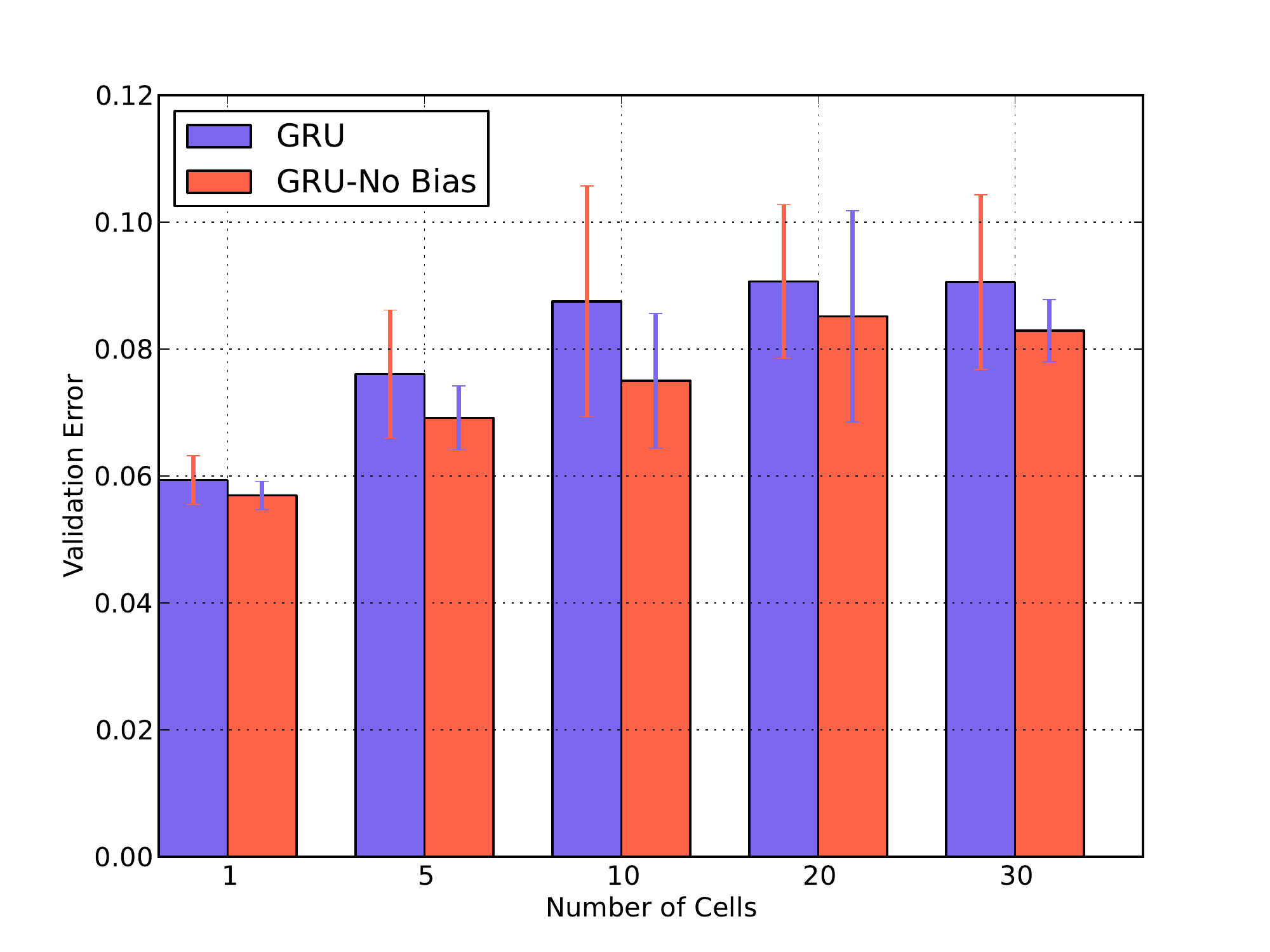}
         \caption{GRU - No Noise.}
	\label{fig:BiasLayersGRU}
    \end{subfigure}
   \begin{subfigure}{.45\linewidth}
	\centering
        \includegraphics[scale=0.4]{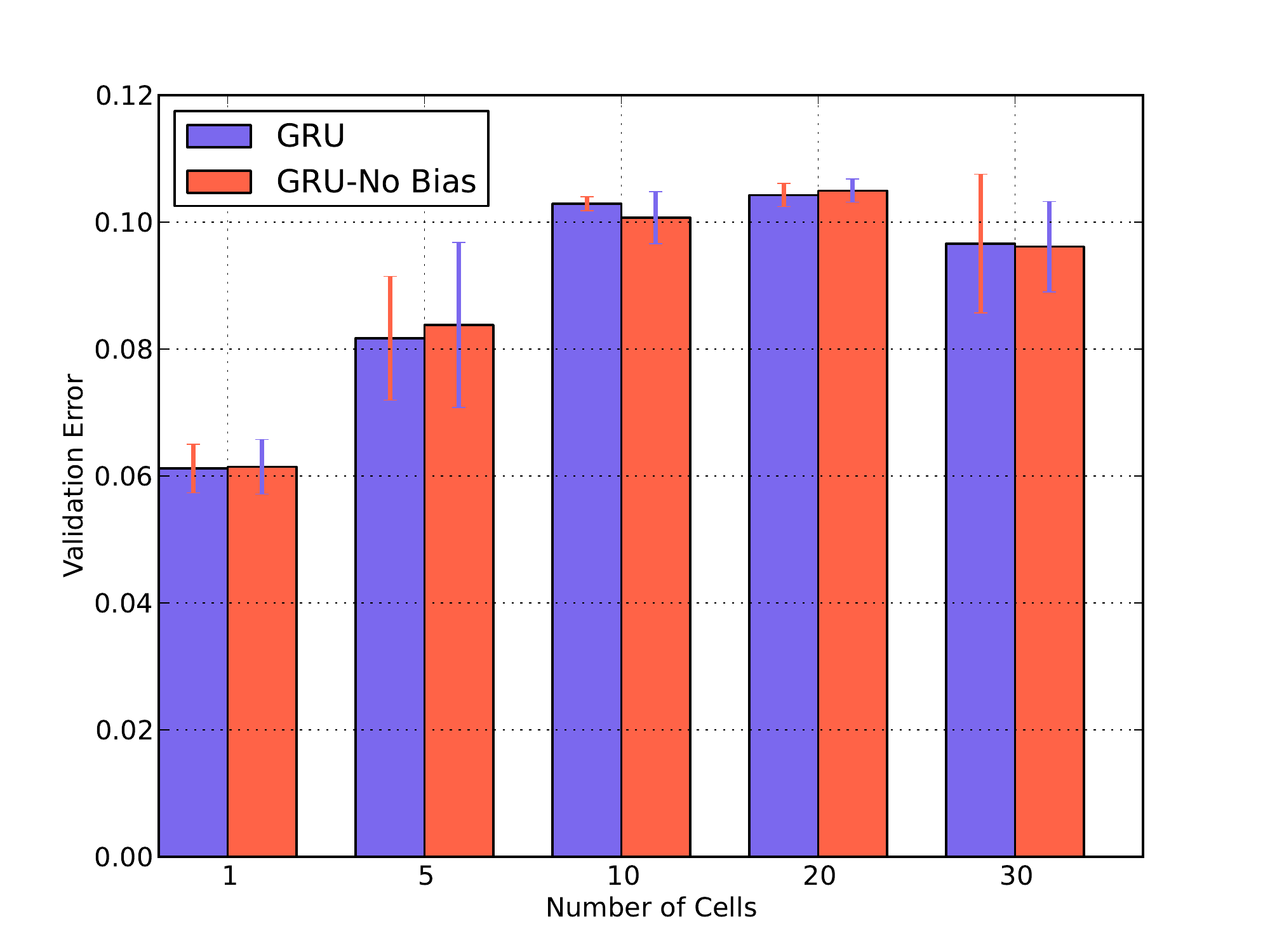}
         \caption{GRU - Washing Noise.}
	\label{fig:BiasLayersGRUWashing}
    \end{subfigure}
  \begin{subfigure}{.45\linewidth}
	\centering
        \includegraphics[scale=0.4]{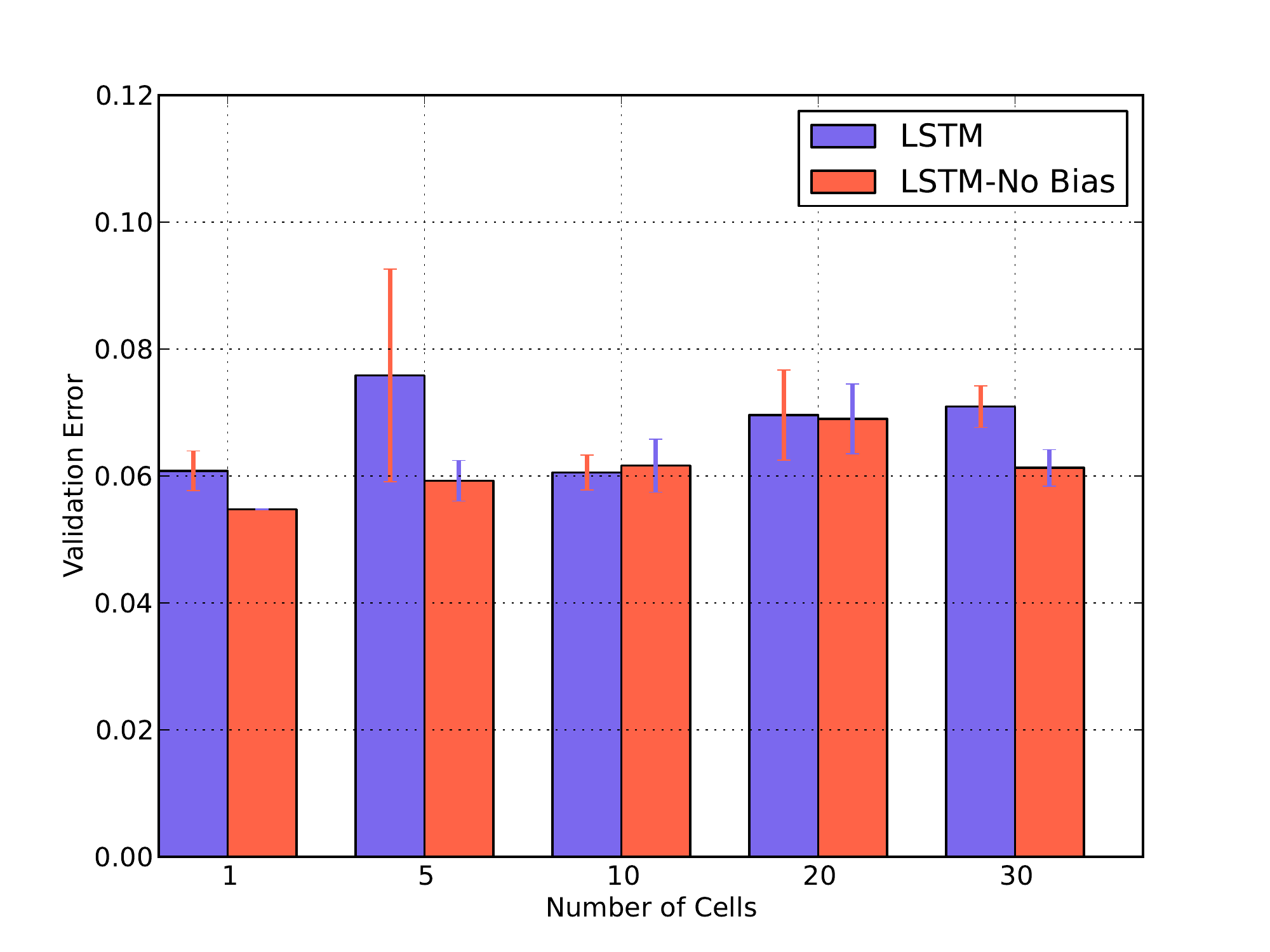}
         \caption{LSTM - No Noise.}
	\label{fig:BiasLayersLSTMRiver}
    \end{subfigure}
    \begin{subfigure}{.45\linewidth}
	\centering
        \includegraphics[scale=0.4]{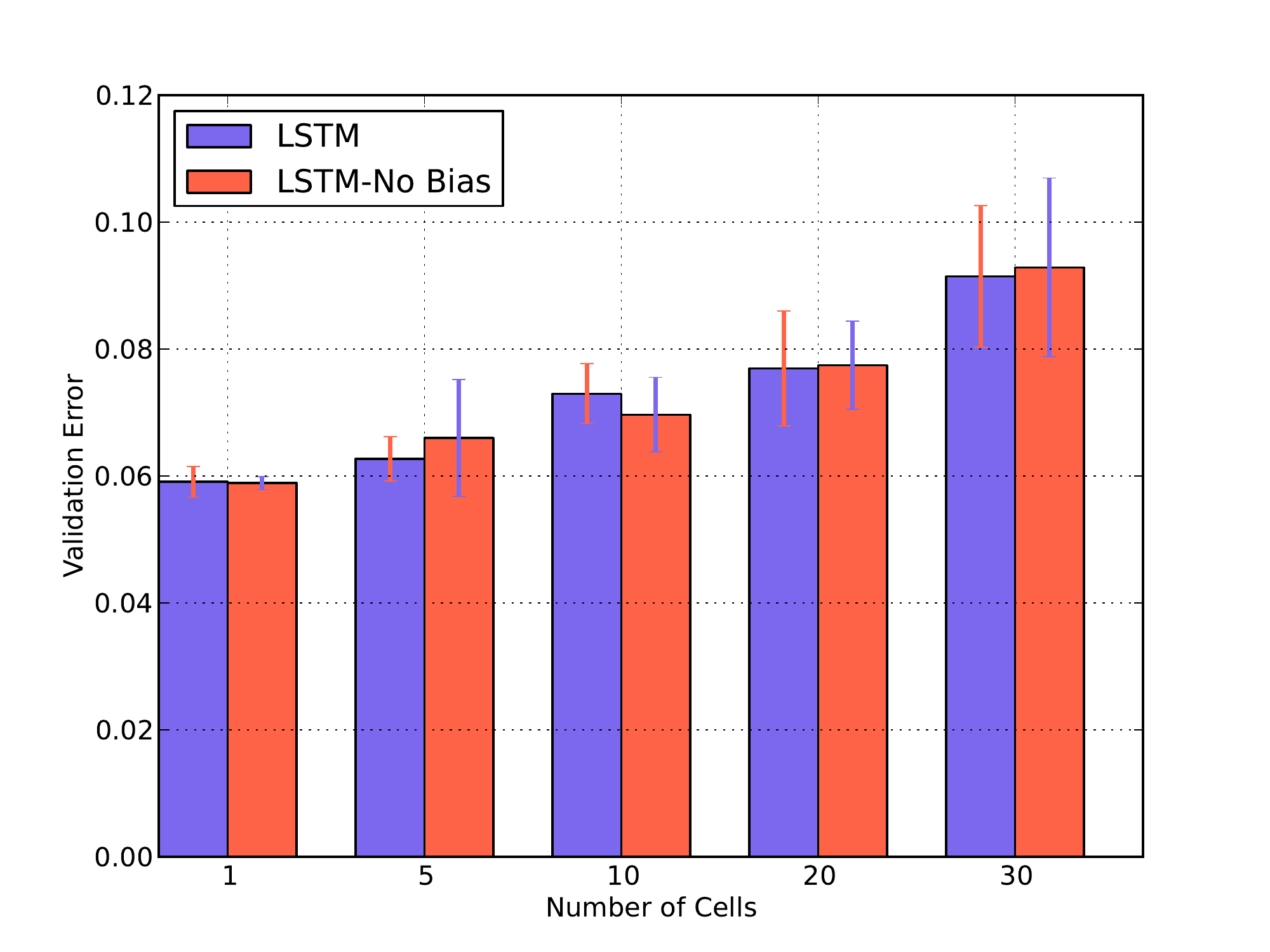}
         \caption{LSTM - Washing Noise.
     }
	\label{fig:BiasLayersLSTMRiver}
    \end{subfigure}
\caption{Classification Performance - Impact of Noise, Bias and Number of Cells.}
\label{fig:CAIBL}
\end{figure*}

\begin{figure*}[ht]
\centering
\begin{subfigure}{.45\linewidth}
\includegraphics[width = 1\linewidth]{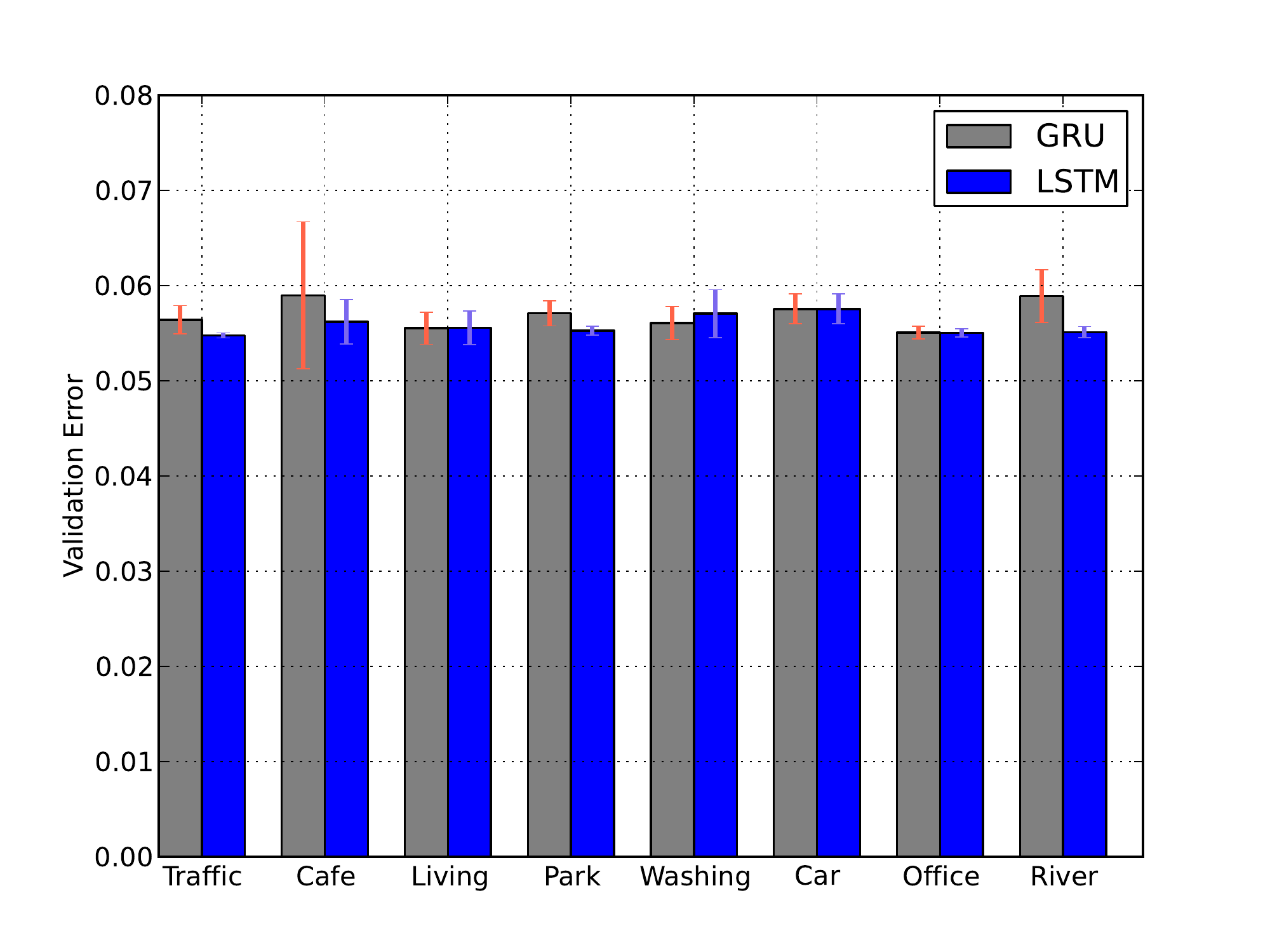}
\caption{}
\label{fig:errorAtNoise}
\end{subfigure}
\begin{subfigure}{.45\linewidth}
\centering
\includegraphics[width = 1\linewidth]{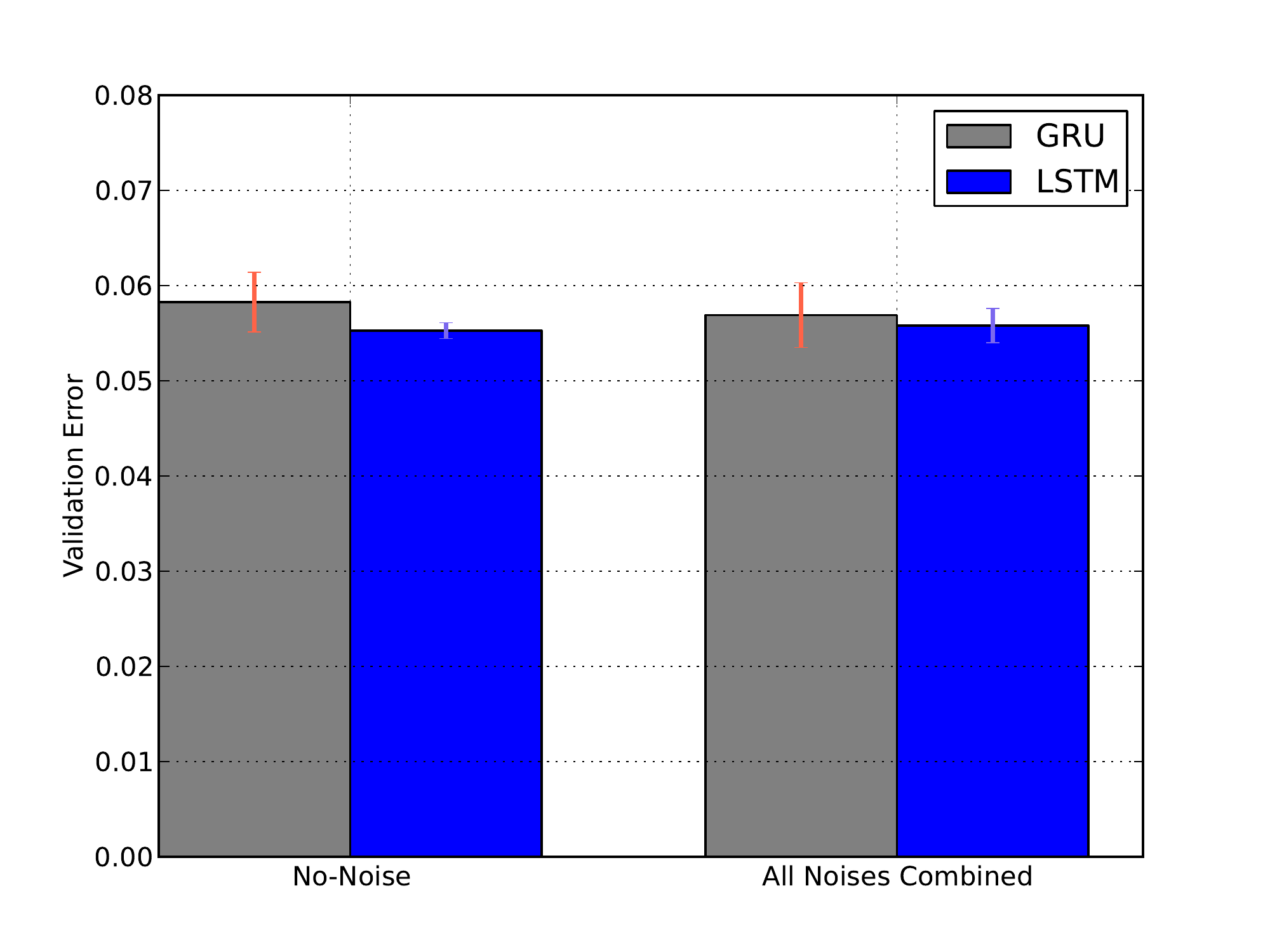}
\caption{}
\label{fig:runTime}
\end{subfigure}
\caption{GRU versus LSTM}
\label{fig:lstmVSGRU}
\end{figure*}
\subsection{Feature Selection}
%The DBN implementation has been adopted from~\cite{keyvanrad2014brief}. The DBN used in the paper uses three RBMs, where the first two RBMs use 1000 hidden unit each, and the third RBM uses 2000 hidden units. 
For each speech segment, we choose a 13 coefficient Mel Frequency Cepstral Coefficients (MFCC) as a feature vector, which have been used by many researchers (e.g., \cite{neiberg2006emotion},\cite{neiberg2006emotion} and many more.) for emotion classification from speech. We choose the small number of coefficients for dimensionality reduction of feature space. Large feature sets consume a significant amount of memory, jointly with computing and power resources and they do not always contribute to improving the recognition rate. 

\section{Results and Discussions}
\label{sec:ER}
 \subsection{Bias and Learning Rate}
We compare the classification performance of GRU with respect to Bias and Learning Rate in Fig.~\ref{fig:cpiNLB}. We reduce the learning rate gradually from 1 to $10^-9$ in the presence and absence of the bias. We find that 
learning rate has a very small impact and GRU performs better without the bias terms. Similar performances have been observed for LSTM. These behaviors can be observed in the noisy conditions as well. We have shown results for arbitrarily chosen Traffic noise.

Setting learning rates for plain Stochastic Gradient Descent in a neural network is usually a process of starting with an initial arbitrary value such as $0.01$ and then doing a cross-validation to find an optimal value. Common values range over a few orders of magnitude from $0.0001$ up to $1$. A small value is desired to avoid overshooting, but a very small value cannot be chosen to avoid getting stuck in local minima or taking long to descend. For our experiments, a learning rate of $1$ yields the best performance. 

Again a Bias value usually allows to shift the activation function to the left or right, which might be necessary for successful learning. However, in our case adding a bias has a negative impact, it lowers the accuracy.

\subsection{Bias and Number of Cells}
We also verify the impact of Number of Cells in presence and absence of Bias in Fig. 5. For both GRU and LSTM, the error is minimum when the number of cells is one and there is no bias term. Similar behavior is observed in presence of noise. We have shown the results for an arbitrary chosen Washing noise.

Both GRU and LSTM layers consist of blocks which in turn consist of cells. We anticipate that due to the smaller size of our dataset the increase in the number of cells does not have a positive impact

\subsection{Accuracy: GRU versus LSTM}
After analyzing the values of Number of Cells and Learning Rate and Bias, we now compare the performance of GRU and LSTM in various noisy conditions in Fig.~\ref{fig:lstmVSGRU}. In the following, we use the following values for the three parameters: \textit{Number of cell = $1$}, \textit{Bias = False}, \textit{Learning Rate = $1$}.

Out of eight different noise cases, in three cases LSTM and GRU performs the same (see Fig.~\ref{fig:errorAtNoise}). While imputed with the Washing  noise, GRU performs better than LSTM by $1.75\%$. For four remaining noise imputations LSTM performs better than GRU. Amongst these, for Cafe and River noise LSTM performs noticeably ($4.6\%$ and $6.4\%$, respectively) better than GRU, but for Traffic and Park noise, it only performs marginally better than GRU. 

The comparison results between LSTM and GRU provide some intuitive insights. For example, GRU performs better for the Washing noise which can be very periodic and not usually continuous. Whereas, LSTM performs noticeably better than GRU in the case of River and Cafe, which are usually sources of continuous noise. We will conduct further studies in future to explain the differences between GRU and LSTM.

Using Fig.~\ref{fig:runTime} we can understand the impact of noise 
(as a whole) on GRU's performance where we compare the error at no-noise with the error combined at all other noisy conditions. We notice that GRU is quite robust to noise. In fact, the error at noisy condition is smaller than the error at the no-noisy condition. It is not unnatural for Deep Learning models to perform better in the presence of noise as this helps avoid overfitting, when the noise is not dominant. We observe similar robustness for LSTM.
\subsection{Run-Time: GRU versus LSTM}
We compare the run-time performance of GRU and LSTM on a 2 GHz Intel Core i7 Macbook professional with a 8 GB 1600 MHz DDR3 memory. For each noise type and for the no-noisy condition we determine the run-time five times. To aggregate over the five times, we choose the median value over mean, as the variances were quite high. We then combine the median values and comparing the averaged median values found that GRU incurs 18.16\% smaller run-time compared to LSTM.

This comparison although were made on desktop Computer, it provides important insights about the time complexity differences between GRU and LSTM. In our future study, we aim to deploy the GRU module on smartphone and determine the run-time complexity.

\section{Existing Work}
\label{sec:EW}
The closest match of our work is the work done by Tang et al.~\cite{tang2016question,tang2016analysis}, where authors use Gated Recurrent Unit for question detection from speech.  Our focus is instead on emotion detection from speech.

 In this paper, we have used LSTM as the benchmark to evaluate the performance of GRU, as LSTM is the most popular Recurrent Neural Networks implementing the gating mechanism.  In particular, we justify the exploration of GRU for emotion classification from speech due to the  success of LSTM in emotion classification. For example, LSTM as a gated recurrent neural network has been shown to successfully exploit the long-range dependencies between successive observations and offer good classification accuracy in ~\cite{wei2014multimodal,wollmer2013lstm}. It has also been shown in the literature that LSTM can be combined with other methods like Multiple Kernel Learning (MKL)~\cite{wei2014multimodal} and Convolutional Neural Networks (CNNs)~\cite{trigeorgis2016adieu} to achieve greater accuracy. It has also been shown that LSTM~\cite{tian2015emotion} can outperform the widely used (e.g.,~\cite{rana2013gait,rana2015simpletrack, rana2015novel}) Support Vector Machine (SVM) when there is enough training data. 

Most of the studies described above consider emotion recognition from clean speech, but we are interested in emotion recognition from noisy speech. One paper by Zhang et al.~\cite{zhang2016facing} performs extensive evaluations of LSTM for speech emotion recognition in presence of non-stationary additive noise and convolutional noise. Whereas these are synthetic noises (additive Gaussian noise), we are more interested in speech emotion recognition in presence of real-life background noise. Also, our focus is mainly on GRU.

\section{Conclusion}
\label{sec:CON}
This paper investigates the feasibility of Gated Recurrent Unit (GRU), a gated Recurrent Neural Network, for emotion classification from noisy speech.  We create noisy speech upon superimposing noises from the cafe, washing, river etc. The results show that GRU offers a very comparable accuracy to the most widely used Long Short-Term Memory (LSTM), while incurring a shorter run-time. For example, LSTM performs better than GRU by 6.4\% in the best case, but GRU incurs 18.6\% smaller run-time compared to LSTM. Interestingly, for washing noise GRU incurs 1.75\% smaller error compared to LSTM. This accuracy versus time-complexity trade-off of GRU is highly advantageous for any embedded platform in general and in our future studies, we aim to investigate the performance of GRU for real-time emotion recognition on smartphones.

\bibliographystyle{IEEEtran}
\bibliography{gruBIB}

\begin{IEEEbiography}
[{\includegraphics[height=1in,keepaspectratio]{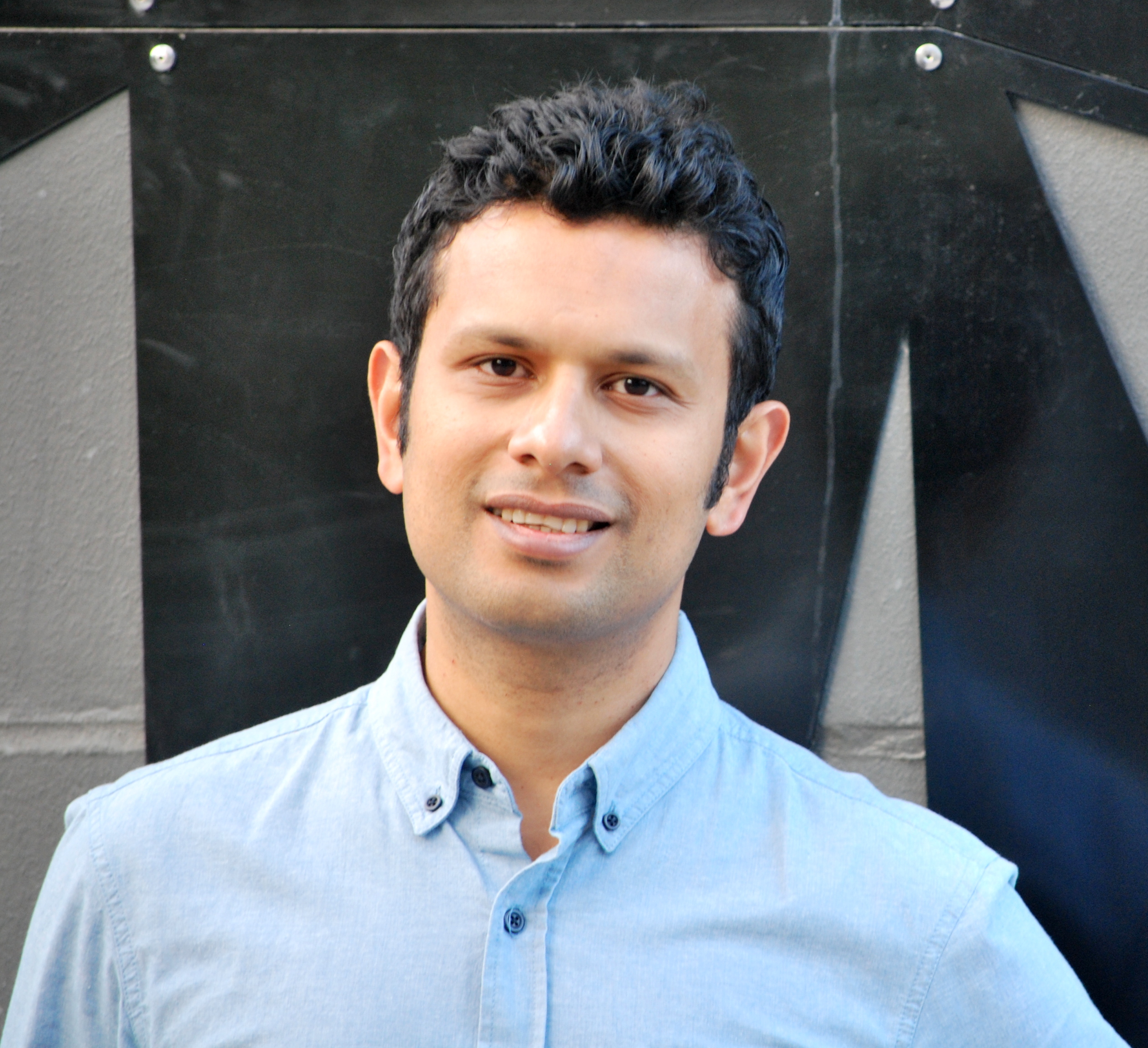}}]{Dr Rajib Rana}
is a Vice Chancellor’s research fellow in the Institute of Resilient Regions, University of Southern Queensland (USQ) and an honorary fellow at the University of Queensland. He received his PhD (2011) in Computer Science and Engineering from University of New South Wales (UNSW), Sydney, Australia. He was the recipient of the President’s and Prime Minister’s Gold Medals for his extraordinary achievement in his Bachelor degree (2004). His current research interests are in Deep Learning Neural Networks, Quantified Self, Personal Health Informatics and the Internet of Things (IoT). He is the founder of the IoT-Health Lab at USQ, where he leads the research on Early Detection of Mental Illness and Autism Spectrum Disorder. Since 2011 Dr Rana has received a number of research grants including, Queensland Health Innovation Grant (twice: 2014, 2016), USQ major infrastructure grant, and UNSW Post-doctoral writing fellowship. He is the lead guest editor of the Special Issue on "Sensors: Deep Learning and Wearable Sensing" (DLWS).

\end{IEEEbiography}

\vspace{-3cm}
\begin{IEEEbiography}
[{\includegraphics[height=.8in,keepaspectratio]{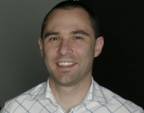}}]{Dr Julien Epps}
is an Associate Professor in Digital Signal Processing with the School of Electrical Engineering and Telecommunications at The University of New South Wales, Sydney, Australia. He also has an appointment as a Scientific Advisor for Boston-based startup Sonde Health, where I work on speech-based assessment of mental health, and have an appointment as a Contributed Principal Researcher with CSIRO, in the ATP Laboratory, where he works on methods for automatic task analysis using behavioural and physiological signals. His research interests also include applications of speech modelling and processing, in particular to emotion and mental state recognition from speech, and genomic sequence processing. He has also worked on aspects of human computer interaction, including multimodal interfaces and computer-supported cooperative work.
\end{IEEEbiography}
\vspace{-2cm}
\begin{IEEEbiography}
[{\includegraphics[height=1.2in,keepaspectratio]{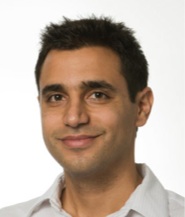}}]{Dr Raja Jurdak} is a Senior Principal Research Scientist at CSIRO, where he leads the Distributed Sensing Systems Group. He has a PhD in Information and Computer Science at University of California, Irvine in 2005, an MS in Computer Networks and Distributed Computing from the Electrical and Computer Engineering Department at UCI (2001), and a BE in Computer and Communications Engineering from the American University of Beirut (2000). His current research interests focus on energy-efficiency and mobility in networks. He has over 100 peer-reviewed journal and conference publications, as well as a book published by Springer in 2007 titled Wireless Ad Hoc and Sensor Networks: A Cross-Layer Design Perspective. He regularly serves on the organizing and technical program committees of international conferences (DCOSS, RTSS, Sensapp, Percomm, EWSN, ICDCS). Dr. Jurdak is an Adjunct Professor at Macquarie University and James Cook University, and Adjunct Associate Professor at the University of Queensland and the University of New South Wales. He is a Senior Member of the IEEE.​
\end{IEEEbiography}

\begin{IEEEbiography}
[{\includegraphics[height=1.2in,keepaspectratio]{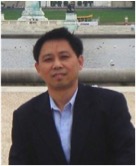}}]{Professor Xue Li} is a Professor in Information Technology and Electrical Engineering at the University of Queensland (UQ) in Brisbane Australia. He has more than 160 research articles published in ACM, IEEE journals and International conferences, books, book chapters, since 1993.  Dr Xue Li was an Editor on Board, (2004-2007) of International Journal of Information Systems and Management \& Technology, Elsevier, A Guest Editor for three issues of the journals: Journal of Global Information Management (JGIM), International Journal of Data Warehouse and Mining (IJDWM), and International Journal of Systems Science (IJSS).  Xue has had more than 29 years’ experience in Information Technology. He is currently a guest professor in three Chinese 985 universities: Central South University of China, Chongqing University, and Xian Jiaotong University. Xue is the founder and Steering Committee Chair of a Southeast Asian International Conference on Advanced Data Mining and Applications 	(ADMA) 2004 2015. He has been invited as a keynote speaker on Big Data Analytics, Data Mining, and Web Information Systems in numerous international conferences.
\end{IEEEbiography}

\vspace{-3cm}
\begin{IEEEbiography}
[{\includegraphics[height=1.2in,keepaspectratio]{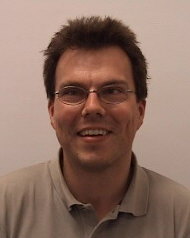}}]{Professor Roland Goecke} 
is a Professor of Affective Computing at the newly created (merged) Faculty of Education, Science, Technology and Mathematics at the University of Canberra. Before that, he was an Associate Professor in Software Engineering at the same Faculty and prior to that an Assistant Professor (Senior Lecturer) at the Faculty of Information Sciences and Engineering, University of Canberra from Janaury 2010 until December 2012. He leads the Vision and Sensing Group and is the Deputy Director of the Human-Centred Computing Laboratory. His research focus continues to be in the areas of face and facial feature tracking and its applications, and more generally in Computer Vision, Affective Computing and Multimodal Human-Computer Interaction. 
\end{IEEEbiography}
\vspace{-3cm}
\begin{IEEEbiography}
[{\includegraphics[height=1in,keepaspectratio]{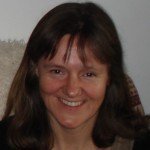}}]{Professor Margot Brereton} researches the participatory interaction design of ubiquitous computing technologies and their interfaces.  She develops innovative designs, methods, and theoretical understandings by designing to support real user communities in selected challenging contexts. Her approach is highly iterative and often involves growing user communities as the design evolves, by understanding and responding to socio-cultural factors.  Her broad area of research include Human-Computer Interaction, Participatory Design, Interaction Design, Computer Supported Cooperative Work, Design Methods, Ubiquitous Computing. \end{IEEEbiography}

\vspace{-3cm}
\begin{IEEEbiography}
[{\includegraphics[height=1.1in,keepaspectratio]{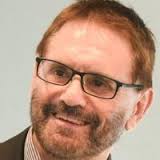}}]{Professor Jeffrey Soar}
holds the Chair in Human-Centred Technology at the University of Southern Queensland where he researches technology innovation for human benefit. He came into research from a career at the highest levels of ICT management within large public and private organisations. He was CIO in several government departments in Australia and New Zealand including Director of Information and Technology for NZ Police where he managed the largest-ever ICT project impacting across emergency services. In academia he established two demonstration smart homes with the latest in innovations for independent living, entertainment, security and energy. His research has been supported by 7 Australian Research Council grants as well as over 30 grants from national and international technology and service organisations. His current research projects are in Technology for Economic Development, E-Learning and M-Learning, E-Government, E-Health, Decision Support, Mobile, Cloud, Algorithms, Adoption and Benefits Realisation, Human Computer Interaction and User Experience.
\end{IEEEbiography}

\end{document}